\documentclass[pra,showpacs,twocolumn,superscriptaddress,amsmath,amssymb]{revtex4-1}

\pdfoutput=1

\usepackage{graphicx}
\usepackage{color}
\usepackage{dcolumn,amsmath}
\usepackage[normalem]{ulem}
\usepackage{xfrac}

\begin{document}

\title{Orbital quantum magnetism  in spin dynamics of strongly interacting magnetic lanthanide atoms} 

\author{Ming Li}
\affiliation{Department of Physics, Temple University, Philadelphia, Pennsylvania 19122, USA}
\author{Eite Tiesinga}
\affiliation{Joint Quantum Institute and Center for Quantum Information and Computer Science,
National Institute of Standards and Technology and the University of Maryland, Maryland 20899, USA}

\author{Svetlana Kotochigova}
\affiliation{Department of Physics, Temple University, Philadelphia, Pennsylvania 19122, USA}
\email[Corresponding author: ]{skotoch@temple.edu}

\begin{abstract} 
Laser cooled lanthanide atoms are ideal candidates with which to study
strong and unconventional quantum magnetism with exotic phases.   
Here, we use state-of-the-art closed-coupling simulations to model
quantum magnetism for pairs of ultracold spin-6
erbium lanthanide atoms placed in a deep optical lattice. In contrast to the
widely used  single-channel Hubbard model description of atoms and molecules
in an optical lattice, we focus on the single-site {\it multi-channel} spin
evolution due to spin-dependent contact,  anisotropic van der Waals, and
dipolar forces.  This has allowed us to identify the leading mechanism, {\it
orbital anisotropy}, that governs molecular spin dynamics among erbium atoms.
The large magnetic moment and combined orbital angular momentum of the
4f-shell electrons are responsible for these strong anisotropic 
interactions and unconventional quantum magnetism.
{\it Multi-channel} simulations of magnetic Cr atoms under similar
trapping conditions show that their spin-evolution is controlled by
spin-dependent contact interactions that are distinct in nature from
the orbital anisotropy in Er. The role of an external magnetic field and the
aspect ratio of the lattice site on spin dynamics is also investigated.

\end{abstract}


\maketitle

Magnetic moments of atoms and molecules originate from their electron's
{\it intrinsic} spin as well as their orbital angular momentum.  In solids
the orbital component of the magnetic dipole moment revolutionized
spintronics research and led to a novel branch of electronics, orbitronics
\cite{Orbitronics2017}.  The orbital magnetic moment is large in materials containing partially
filled inner shell atoms.  The distinquishing feature of such atoms
is the extremely large anisotropy in their interactions.  This orbital
anisotropy is the crucial element in various scientific applications and
magnetic technology \cite{Bernevig2005,Kontani2008,Scagnoli2011}. 

Inspired by the role of orbital anisotropy in magnetic solids and to
deepen our knowledge of quantum magnetism, we  study orbital anisotropy
at the elementary level by capturing the behavior of strongly-interacting
magnetic lanthanide atoms in an optical lattice site. In particular, we
simulate the time-dependent {\it multi-channel} spin-exchange dynamics of
pairs of interacting erbium atoms in sites of a deep three-dimensional
optical lattice with negligible atomic tunneling between lattice
sites.

Theoretical spin models have a long established role as useful tools for
understanding interactions in magnetic materials.  
Recently, researchers demonstrated that laser-cooled ultracold atoms and
molecules in optical lattices are a nearly perfect realization of these
models with control of the local spin state and spin-coupling strength
\cite{Bloch2005,Barnet2006,Rempe2007,Trotzky2008,Wall2009,Wall2010,Daley2010,Krems2010,Gorshkov2011,Ray2014,Ferlaino2016}.
However, most of the microscopic implementation of spin models has focussed on atoms 
with zero orbital angular momentum and often based on a single-state (Fermi- or Bose-) Hubbard model description \cite{Jaksch1998,Sanpera2007}. 

The limitation of simplified Hubbard models essentially holds true for
optical lattices filled with magnetic lanthanide atoms that have open
electronic 4f-shells and possess a large unquenched orbital angular
momentum. These atoms are now emerging as a promising platform for the
investigation of high-spin quantum magnetism.  Theoretical simulations
of ultracold lanthanide atoms in an optical lattice is a difficult task
that requires multi-channel analyses of interactions and correlations.

Although quantum many-body effects have been
observed and studied with quantum gases of lanthanide atoms
\cite{Pfau2016,Ferlaino2016,Burdick2016,Schmitt2016,Chomaz2018}, the on-lattice-site
spin dynamics of highly magnetic atoms remains not fully understood as
it requires a precise knowledge of the two-body interactions.

We, first, focus on single-site {\it multi-channel} spin evolution due to
spin-dependent contact, anisotropic van-der-Waals, and dipolar forces.
Then, we go beyond the one-site model in order to
account for coupling between our atom pair and other (doubly-)occupied
lattice sites and find a slow damping of the local spin oscillations.
In our Mott insulator regime, the single-site quasi-molecule is
surrounded by atom pairs in neighboring sites. This only permits the
magnetic dipole-dipole coupling between the atom pairs justified by
studies \cite{Hensler2003,Trotzky2008} showing that for short-range
 interactions between atoms in neighboring lattice sites is
orders of magnitude weaker than the  dipole-dipole interaction.

Dipole-dipole interaction induced quantum magnetism was pioneered in
experiments of Laburthe-Tolra's group ~\cite{Laburthe2013}. The authors
used  magnetic chromium atoms with their three units of angular momenta
and demonstrated the most-dramatic spin oscillations observed to date.
These oscillations resulted from the strong coupling between the spins,
both from  isotropic  spin-dependent on-site contact interactions
and between-site magnetic dipole-dipole interactions.

We have shown here that the orbital-induced molecular anisotropy, absent in
alkali-metal and chromium collisions, is much stronger in interactions of lanthanide atoms
than that of the magnetic dipole-dipole interaction at interatomic separations smaller
than $200a_0$, where $a_0$ is the Bohr radius. At shorter separations
the strength of this anisotropy for Dy and Er atoms is about
10\% that of the spin-independent isotropic interaction.

Following Refs.~\cite{Frisch2014,Maier2015,Baier2016},
we prepare  ground-state spin-6 erbium atom pairs in the energetically-lowest
vibrational state of a lattice site and  
Zeeman sublevel $|j_a, m_{a}\rangle = |6,-6\rangle$,
where $j_a$ and $m_{a}$ are the  atomic angular momentum  and its projection
along an applied magnetic field, respectively. To initiate spin dynamics
we  transfer each atom to  Zeeman sublevel $|6,-5\rangle$ with a short rf
pulse. We single out this   pair state among the many internal
hyperfine  states, due to its ability to collisionally evolve
into the $|6,-6\rangle + |6,-4\rangle$ molecular state. We then use
the spin-oscillation frequency of the atomic populations to extract the
energy splitting between $|6,-5\rangle + |6,-5\rangle$ and $|6,-6\rangle +
|6,-4\rangle$. This magnetic-field dependent splitting is, unlike for alkali-metal and chromium
atoms, due to strong orbital anisotropy in interactions between the atoms. 

Details of the Hamiltonian of an Er atom pair in a lattice
site modeled by a cylindrically-symmetric harmonic potential,
the computational tools, analysis of the
eigenpairs, and state preparation by rf pulse can be found in the Supplemental
Material. For later use we define the ``€˜isotropic'' trap frequency $\omega$
via $\omega^2=(\omega_z^2+2\omega_\rho^2)/3$, where $\omega_z$ and
$\omega_\rho$ are the axial and radial harmonic frequencies, respectively.
Eigenstates $|i,M\rangle$ with $B$-dependent energy $E_{i,M}$ are uniquely
labeled by  projection $M$ of the total molecular angular momentum
and integer index $i$.

\begin{figure}
\includegraphics[scale=0.3, trim=0 0 0 0,clip]{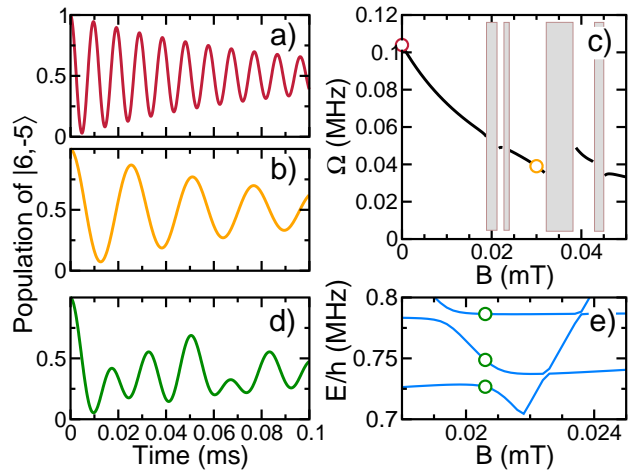}
\caption{Spin population dynamics of $^{168}$Er in an isotropic trap with $\omega/(2\pi)$ = 0.4 MHz. 
Panels a) and b) show  sinusoidal time
traces of the population in $|6,-5\rangle$ for  $B=0.1$ $\mu$T and
0.03 mT, respectively.  Panel c) shows the spin-oscillation frequency as
a function of $B$ with blocked out field regions (gray bands) where the
oscillations are not sinusoidal. The two open circles correspond to
the field values shown in panels a) and b). A complex oscillation pattern
for $B=0.0206$ mT, located in one of the banded regions, is shown in panel d) .  Panel e)
shows the corresponding  avoided crossing between the three populated  
energy levels.  We  assume a slow damping
rate of $\gamma=1.2\cdot 10^4$ s$^{-1}$. Its
origin is discussed in the text.}
\label{PopDynamics}
\end{figure}

In our model, the molecular interactions $U$,  i.e.  operators that depend on separation $r$, contain
a dominant angular-momentum-independent potential with strength $V_0(r)$,
an isotropic spin-spin interaction with strength ${V^{\rm jj}(r)\propto 1/r^6}$ that entangles the spins
of the  atoms, and anisotropic terms that entangle spin
and orbital degrees of freedom. The latter terms are separated into a magnetic dipole-dipole
and orbital contribution with  strengths ${V^{\rm dip}(r)\propto 1/r^3}$ and ${V^{\rm orb}(r)\propto 1/r^6}$, respectively.
These interactions play a similar role as spin-orbit
couplings induced by synthetic magnetic fields \cite{Lin2009,LeBlanc2012}.
The short- and long-range shapes of of $U$ are determined from a combination of
experimental measurements and {\it ab~initio} electronic-structure
calculations \cite{Kotochigova2014,Maier2015}.   Our results, unless
otherwise noted, are based on an $U$ that
reproduces the experimentally-determined scattering length of $137a_0$ for
the $^{168}$Er $|6,-6\rangle$ Zeeman level near zero magnetic field \cite{Baier2016}. 
Consequences of the uncertainties in the potentials will be discussed
later on.

Figures~\ref{PopDynamics}a), b), and d) show our predicted $|6,-5\rangle$
population as a function of time after the short rf pulse for three characteristic, but small $B$. The
population is seen to oscillate.  Those in panels a) and b) are
single sinusoids with a frequency $\Omega(B)$
and reflect the fact that only two  $|i,M=-10\rangle$ are populated by the pulse. 
We find that these states to good approximation are labeled by diabatic states $|\ell; j,m\rangle\rangle=
| s; 10,{-10}\rangle\rangle $ and $|s; 12,{-10}\rangle\rangle$,
where the $s$ corresponds to $\ell=0$, $s$-wave partial wave scattering
and $\vec j$ is the sum of the two atomic angular momenta ($m$ is its projection).
Figure~\ref{PopDynamics}c) shows that the frequency $\Omega$ is a sharply decreasing function of $B$.

In the banded regions  of Fig.~\ref{PopDynamics}c) avoided crossings between three states occur 
and the spin oscillations have multiple periodicities. An example from near $B=0.02$ mT is shown 
in panel d) together with a blowup of the relevant populated energy levels in panel e).   
The third state away from this avoided crossing is characterized by $|d; 12,-12\rangle\rangle$
with an energy that decreases with $B$ and a large $d$-wave character.
Near the avoided crossing  eigenstates are superpositions of the three diabatic
states $|k\rangle\rangle=| s; 10,{-10}\rangle\rangle $, $|s; 12,{-10}\rangle\rangle$ and $|d; 12,-12\rangle\rangle$.
See Supplemental Material for more details.
Neither  the $B$-dependence of $\Omega(B)$  nor
the presence of avoided crossings can be observed  in atomic chromium and
both are solely the consequence of the anisotropic dispersive interactions
in magnetic lanthanides and form two important results of this article.

In an experiment   thousands of simultaneous spin-oscillations
occur, one in each site of a $D$-dimensional optical lattice.
Atom pairs in {\it different} lattices sites are  coupled
by magnetic dipole-dipole interactions. This leads to dephasing
of the intra-site spin oscillation. Here, we  estimate the timescale
involved.  The inter-site dipolar strength for a typical lattice period
$\delta\lambda$ between 250 nm and 500 nm is an order of magnitude smaller
than the energy spacings between the local   $|s; j,m_j\rangle\rangle$ and $|d;
j,m_j\rangle\rangle$ states.  In principle, this
dipole-dipole interaction  can change  di-atomic projection
quantum numbers $j$ and $m_j$.  We, however, focus on  couplings that are
superelastic and ignore exchange processes. That is, transitions that
leave  $j$ and the sum of the Zeeman energies unchanged, i.e. $|s; j_p
m_p\rangle\rangle_p |s; j_q m_q\rangle\rangle_q \longleftrightarrow |s;
j_p {m_p+1}\rangle\rangle_p|s; j_q {m_q-1}\rangle\rangle_q$ etc., where
the subscripts $p$ and $q$ on the kets indicate different  lattice
sites.  For $N$ unit cells  and focusing on states $|s;
j,m_j\rangle\rangle_p$ of a single $j$ this leads to $\approx(2j+1)^N$
spin configurations when $N\gg 2j+1$ and with nonuniformly-distributed
eigenenergies that span an energy interval of order $2D\times\mu_0/(4\pi)
\times(g\mu_B j)^2/\delta\lambda^3\equiv\Delta$   accounting for nearest-neighbor coupling only.
Here, $g$ is the atomic g-factor, $\mu_B$ is the Bohr magneton, and $\mu_0$ is the magnetic constant.
For  Er$_2$, $\Delta/h=D\times 320$ Hz  with $j=12$ and
$\delta\lambda=250$ nm and Planck constant $h$.  The value $\Delta$ is  a lower bound for
the energy span.

We then simulate the intra-site spin-evolution in the presence of
these super-elastic dipolar processes with di-atoms in other sites 
with a master equation for the reduced density matrix  $\rho_{kk'}(t)$
with up to three di-atomic basis states $|k\rangle\rangle$ and Lindblad
operators $L_k=\sqrt{\gamma_k}|k\rangle\rangle\langle\langle k|$ \cite{Cohen2004}
that damp coherences but not populations at rate $\gamma_k=2\pi\eta_k
\Delta/h$ with dimensionless parameters $\eta_k$ of the order
unity. Figures \ref{PopDynamics}a), b), and d)
show this dephasing assuming $\gamma_k=\gamma=1.2\cdot 10^4$ s$^{-1}$ for
all $k$ (i. e. $\eta_k=2$ and $D=3$).  For $t\to\infty$ the  overlap
with the initial state approaches $|c_{j=10}|^4+|c_{j=12}|^4\approx
0.50$ for panels a) and b) and to a value that depends on the
precise mixing of the three diabatic states  for panel d).  The $c_j$ are defined in the
Supplemental Material.

The short-range shapes of the Er interaction potentials have significant uncertainties
even when taking into account of the $137a_0$ scattering length of the $|6,-6\rangle$ state.
This modifies the expected spin-oscillation frequencies.  We characterize
the distribution of the spin-oscillation frequency by changing the depth
of the isotropic and spin-independent $V_0(r)$ over a small range, such
that its number of $s$-wave bound states changes by one, while keeping its
long-range dispersion coefficient fixed. The nominal number of bound states is $72$,
much larger than one, and based on quantum-defect theory \cite{Gribakin1993}
we can  assume that each depth within this range is equally likely.

\begin{figure}
\includegraphics[scale=0.2,trim=0 15 0 0,clip]{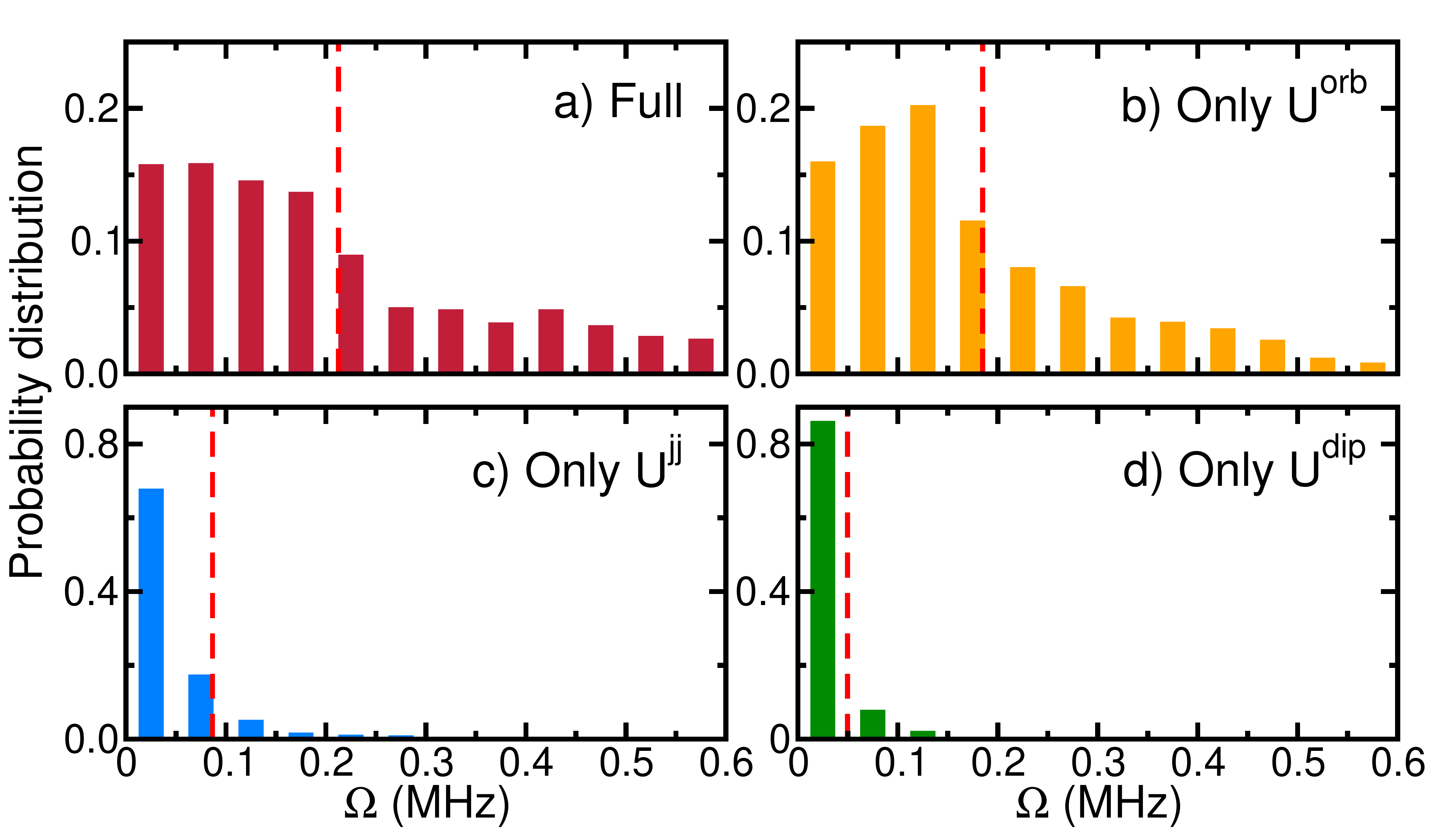}
\caption{Probability distributions of the ${M=-10}$ spin-oscillation
frequency $\Omega$ showing the role of anisotropic  interactions
in $^{168}$Er.  Data is for an isotropic trap with $\omega/ 2\pi = 0.4$
MHz and $B=0.1$ $\mu$T.  Panels a), b),  c), and d) show distributions
for the full interaction potential $U$,  $U\to V_0(r)+V^{\rm orb}(\vec
r)$, $U\to V_0(r)+V^{\rm jj}(r)$, and $U\to V_0(r)+V^{\rm dip}(\vec r)$,
respectively.  In each panel  the mean  spin-oscillation frequency is
indicated by a vertical dashed line.}
\label{Distribution}
\end{figure}

Figure \ref{Distribution}a) shows the distribution of oscillation
frequencies $\Omega=|E_{i,M}-E_{i',M}|/h$ at $B=0.1$ $\mu$T away from avoided crossings assuming  an isotropic
harmonic trap. Here,  the eigenstates are $|i,{M=-10}\rangle\approx | s; 10,{-10}\rangle\rangle$
and $|i',{M=-10}\rangle\approx|s; 12,{-10}\rangle\rangle$.  (Few of the realizations show evidence
of mixing with other states.) The distribution is broad and peaked at
zero frequency.  Its mean frequency is $\approx 0.5\hbar\omega$ and should
be compared to the $2\hbar\omega$ spacing between the harmonic levels.

We have also computed the distribution for Hamiltonians, where one or
more parts of  $U$ have been turned off. In
particular, Fig.~\ref{Distribution}b)  shows the distribution
for the case where $U$ is replaced by  $V_0(r)$ and the term proportional to $V^{\rm orb}(\vec r)$, 
while Figs.~\ref{Distribution}c) and d) show that for $U\to V_0(r)+V^{\rm jj}(r)$ 
and $U\to V_0(r)+V^{\rm dip}(\vec r)$, respectively.  The distribution
in panel b) is about as broad as that in panel a) indicating that
the anisotropic dispersion of $U^{\rm orb}(\vec r)$ is the most
important factor in determining the distribution for the full $U$,
even though the precise distribution has noticeably changed.  In panel
b) smaller and larger splittings are now suppressed relative to those near the mean.
In panel c) with only isotropic spin-spin interactions and panel d)
with only the magnetic dipole-dipole interaction coupling  spins
and  orbital angular momenta the distributions are highly localized.

One of the most-studied ultracold magnetic atom is  chromium
with its magnetic moment of $6\mu_{\rm B}$
and  spin $s=3$ \cite{Pfau2005,Tiesinga2005,Lahaye2009,Stuhler2010,Laburthe2013,Laburthe2016}. This  moment is only
slightly smaller than that of Er, $\approx 7\mu_{\rm B}$. The dipolar
parameter $\epsilon_{\rm dd}=a_{\rm dd}/a$ for these atoms, however, is
very different. Here, dipolar length   
$a_{\rm dd}=(1/3)\times 2\mu (g\mu_{\rm B}j)^2\mu_0/(4\pi\hbar^2)$, $a$ is a relevant  $s$-wave scattering length, and $\mu$ is the reduced mass. In
fact, $\epsilon_{\rm dd}$ is 0.16 for $^{52}$Cr \cite{Lahaye2009} and near two
for $^{168}$Er \cite{Ferlaino2016} for the $|6,-6\rangle$ Zeeman
level due to the large difference in their mass. Another distinction
is their orbital electronic structure.   Chromium has  a $^7$S$_3$ ground
state and no orbital  anisotropy, whereas erbium has a $^3$H$_6$
ground state and a large orbital anisotropy.

Quantum magnetism of pairs of Cr atoms in a lattice site was
investigated in Refs.~\cite{Laburthe2013,Laburthe2016}.  Cr atoms were prepared
in the $|s,m_s\rangle=|3,-3\rangle$  state
and  spin dynamics was initiated by  transfer into  the $|3,-2\rangle$
state. Then the quasi-molecular state $|3,-2\rangle+|3,-2\rangle$ evolves
into  $|3,-3\rangle+|3,-1\rangle$ state and back. The energy difference
between these states is due to molecular interactions.

The  potential operator $U$ for Cr$_2$ only contains 
isotropic interactions except for the magnetic dipole-dipole interaction. These
isotropic potentials  can  be represented as
a sum of tensor operators, $U^{\rm iso}(r)=V_0(r)+
V_{1}(r)\vec s_a\cdot \vec s_b + V_2(r) \,T_2(\vec s_a,\vec s_a)\cdot
T_2(\vec s_b,\vec s_b) +\cdots$ for atoms $a$ and $b$, where $V_0(r)$ has an attractive well
and a dispersion potential with $C_6 = 733 E_{\rm h}a_0^6$  for $r\to\infty$
\cite{Tiesinga2005} and $V_{1}(r)$  is the exchange
potential proportional to $r^\gamma e^{-\kappa r}$ for $r\to\infty$.
The non-negligible $V_2(r)$ is the strength of a second spin-dependent
contribution. It and any other additional operator also decrease
exponentially with $r$.  The rank-2 tensor $T_2(\vec s_a,\vec
s_b)$ is   constructed from angular momenta $\vec
s_{a}$ and $\vec s_{b}$.

\begin{figure}
\includegraphics[scale=0.33,trim=25 17 0 30,clip]{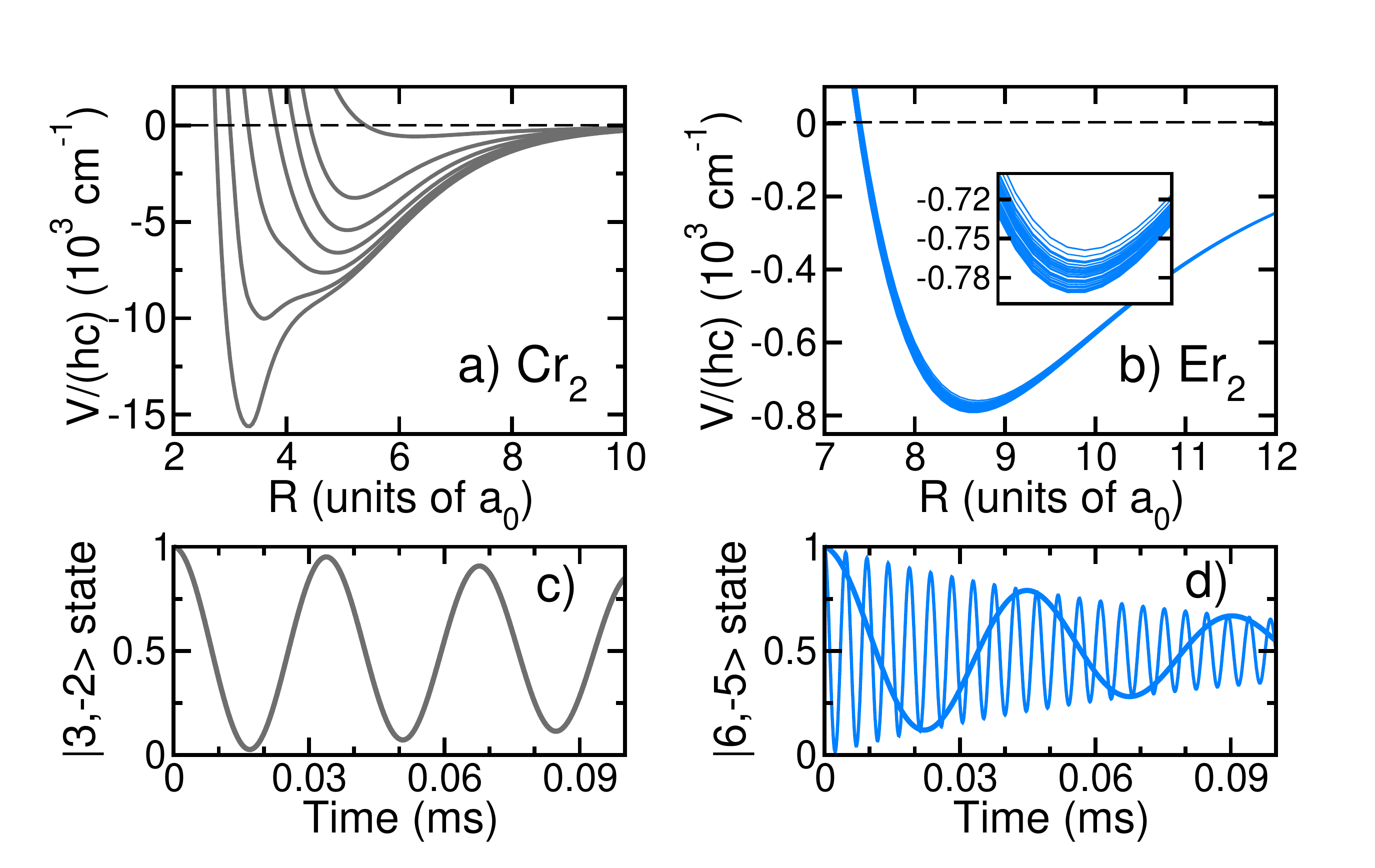}
\caption{Ground-state adiabatic potentials of Cr$_2$ and Er$_2$ and
a comparison of their spin evolution. Panel a) shows the seven $^{2S+1}\Sigma^+$ BO potentials of Cr$_2$ as functions of separation $r$
calculated in Ref.~\cite{Krems2005}.  Panel b) shows the forty-nine
gerade potentials of Er$_2$ with  $\Omega=0,\dots, 12$. Panels c) and d)
show the population evolution of the $|3,-2\rangle$ and $|6,-5\rangle$
states of Cr and Er atoms at $B=0.1$ $\mu$T, respectively. The solid and dashed curves
in panel d) are for the full interaction potential and a potential that only  includes the isotropic
interactions, respectively.  Damping  is
due to dephasing from dipolar interactions with atoms in neighboring
lattice sites. The lattice geometry is the same for both atomic species and as in Fig.~\ref{PopDynamics}.}
\label{Er_Cr} 
\end{figure}

For Cr$_2$ the tensor description of $U$ is equivalent to a description
in terms of Born-Oppenheimer (BO) or adiabatic potentials, labeled
$^{2S+1}\Sigma^+_{g/u}$ with $\vec S=\vec s_a+\vec s_b$ and $S=0,1,\cdots,
6$ (Even and odd spin $S$ corresponds to  {\it gerade}(g) and {\it ungerade}(u)
 symmetry, respectively.)  They were computed in Ref.~\cite{Krems2005} and 
reproduced in Fig.~\ref{Er_Cr}a).  The complex relationship between the BO potentials
indicates that $V_{q>0}(r)$ are on the order of $V_q/(hc)=10^{3}$
cm$^{-1}$ for $r<6a_0$.  Figure \ref{Er_Cr}b) shows the equivalent
graph for two Er atoms as obtained by diagonalizing our $U$ at each $r$
with $\vec r$ aligned along the internuclear axis. The potentials
have depths $D/(hc)$ between 750 cm$^{-1}$ and 790 cm$^{-1}$ at the
equilibrium separation, where the splittings are mainly due to the anisotropic
interaction proportional to $V^{\rm orb}(r)$.  Visually, the adiabatic
potentials of the two species are very distinct. The crucial
physical distinction, however, lies in the origin of the splittings
between the potentials, i.e. isotropic versus anisotropic interactions.

We find it convenient to simulate the spin dynamics of Cr by replacing the  $V_q(r)$ by delta-function or contact potentials
$4\pi(\hbar^2/2\mu)\, {\cal A}_q\, \delta(\vec r)\partial/\partial r$ for $q=0,1$ and $2$,
with fitted lengths ${\cal A}_q$  such that the contact model reproduces
the measured scattering lengths of the BO potentials.
We used ${\cal A}_0=60.6a_0$, ${\cal A}_1=6.73a_0$ and
${\cal A}_2=-0.243a_0$ leading to the measured scattering lengths
of $-7(20)a_0$, $58(6)a_0$, and $112(14)a_0$ for the $S=2$, $4$, and $6$
 potentials, respectively \cite{Tiesinga2005}. (The
numbers in parenthesis are the quoted uncertainties. The scattering
length for the $^1\Sigma_g^+$ potential is not known.
The  ${\cal A}_q$ will change once this  scattering
length is determined.)

Figures~\ref{Er_Cr}c) and d) compare the spin oscillations for a pair of
Cr and Er atoms for the same lattice geometry and $B=0.1$ $\mu$T, respectively.  
The eigen energies of a Cr pair in an isotropic harmonic trap interacting via the delta function
potentials are found with the help of the non-perturbative analytical
solutions obtained in Ref.~\cite{Busch1998} .  The spin evolution for Cr
is solely due to the isotropic spin-spin interactions proportional to
lengths ${\cal A}_{1}$ and ${\cal A}_2$ and are independent of $B$.  
Two time traces for Er  are shown
corresponding to the full $U$ and one where  
only the isotropic interactions are included. 
We see that the oscillation period is  slower when the anisotropic interactions are excluded
and the behavior is much more alike to that of a Cr pair.
The curves in panels c) and d)  also include
an estimate of dephasing due to atom pairs in neighboring lattice sites
again using a lattice spacing of 250 nm. Dephasing of a chromium-pair
 is only 25\% smaller than that of an erbium-pair due to
the slightly-smaller magnetic moment of Cr.

We have presented simulations that give valuable insight into the
interactions of lanthanide-based quantum systems with periodic arrays
or lattices containing atom pairs in each lattice site. As our atoms
carry a spin that is much larger than that of spin-1/2 electrons
in magnetic solid-state materials, this might result in a wealth of
novel quantum magnetic phases. The ultimate goal of any Atomic, Molecular and
Optical implementation of quantum magnetism is to develop a controllable
environment in which to simulate materials with new and advanced
functionality.

In spite of the successes of previous analyses of quantum simulations with
(magnetic) atoms in optical lattices \cite{Baranov_Review2012}, simplified
representations with atoms as point particles and point dipoles can not always
be applied to magnetic lanthanide atoms.  Important information about
the electron orbital structure within the constituent atoms is lost.

In the current study the orbital anisotropy of magnetic-lanthanide
electron configurations is properly treated in the interactions
between Er atoms in optical lattice sites and is used to describe and
predict spin-oscillation dynamics.  We illuminated the role of orbital
anisotropy in this spin dynamics as well as studied the interplay between
the molecular anisotropy and the geometry of the lattice site potential.
The interactions lift the energy-degeneracies of different spin orientations,
which, in turn, leads to  spin oscillations.

\section{Acknowledgments} Work at Temple University is supported
by the AFOSR Grant No. FA9550-14-1-0321, the ARO-MURI Grants
No. W911NF-14-1-0378, ARO Grant No. W911NF-17-1-0563, and the NSF
Grant No. PHY-1619788.  The work at the JQI is supported by NSF Grant
No. PHY-1506343.

\bibliography{Refs}

\end{document}


\title{Orbital Quantum Magnetism in Spin Dynamics of Strongly Interacting Magnetic Lanthanide Atoms: Supplemental Material}

\author{Ming Li}
\affiliation{Department of Physics, Temple University, Philadelphia, Pennsylvania 19122, USA}
\author{Eite Tiesinga}
\affiliation{Joint Quantum Institute and Center for Quantum Information and Computer Science,
National Institute of Standards and Technology and the University of Maryland, 
Gaithersburg, Maryland 20899, USA}
\author{Svetlana Kotochigova}
\affiliation{Department of Physics, Temple University, Philadelphia, Pennsylvania 19122, USA}

\maketitle

\subsection*{Relative Motion of two bosonic lanthanide atoms}

A single site of an optical lattice potential  is well approximated
by a cylindrically-symmetric  harmonic trap.  The  Hamiltonian
for the relative motion of two ground-state bosonic lanthanide atoms
is then $H_{\rm rel} =H_0+U$, where
\[
        H_0 ={\vec p\,}^2\!/(2\mu)+ \mu (\omega_\rho^2\rho^2 +\omega_z^2 z^2)/2 + H_{\rm Z}    
\]
with relative momentum $\vec p$,  relative coordinate $\vec
r=(r,\theta,\varphi)=(\rho, \theta, z)$  between the atoms in spherical
and cylindrical coordinates, respectively, and reduced mass $\mu$.
Moreover,  $\omega_\rho$ and $\omega_z$ are trapping frequencies and
$H_{\rm Z}=g\mu_{\rm B} {(\vec \jmath_a+\vec \jmath_b)\cdot \vec B}$
is the Zeeman Hamiltonian  with atomic g-factor $g$ and Bohr magneton
$\mu_{\rm B}$.  Zeeman states  $|j_a, m_a\rangle|j_b, m_b\rangle$ are
eigen states of $H_{\rm Z}$, where $\vec\jmath_\alpha$ is the total
atomic angular momentum of atom $\alpha=a$ or $b$ and $m_\alpha$ is its
projection along $\vec B$.

We have assumed that the harmonic lattice-site potential is the same
for all atomic states and a  $\vec B$ field directed  along
the axial or $z$ direction of the  trap.  Small tensor light shifts, proportional
to $(m_a)^2$ \cite{Laburthe2013}, induced by the  lattice lasers,  have been omitted.
Here and throughout, we use dimensionless angular momenta, i.e.  $\vec
\jmath/\hbar$ is implied when we write $\vec \jmath$ with $\hbar=h/2\pi$
and Planck constant $h$. For angular momentum algebra we
follow Ref.~\cite{BrinkSatchler}.

The Hamiltonian term $U$ describes the molecular
interactions. It contains an isotropic contribution $U^{\rm iso}( r)$ that 
only depends on $r$ as well as an anisotropic contribution $U^{\rm aniso}(\vec r)$ that also depends on the
orientation of the internuclear axis. In fact, we have
$U^{\rm iso}(r) =  V_0(r) + V^{\rm jj}(r) \, {\vec\jmath_a\cdot \vec \jmath_b} + \cdots$,
where  for ${r\to\infty}$ the angular-momentum independent ${V_0(r)\to-C_6/r^6}$
with  dispersion coefficient ${C_6 =1723E_{\rm h} a^6_0}$ for Er$_2$  \cite{Frisch2014,Maier2015}. 
For small $r$ the potential $V_0(r)$ has a repulsive wall and an attractive well with depth  ${D_e/(hc)\approx790}$ cm$^{-1}$.
We use $V^{\rm jj}(r)= -c^{\rm jj}_6/r^6$ for all $r$  with dispersion coefficient $c^{\rm jj}_6=-0.1718 E_{\rm h} a^6_0$.
Furthermore, $U^{\rm aniso}(\vec r)=U^{\rm orb}(\vec r)+U^{\rm dip}(\vec r)$ with
\begin{eqnarray}
U^{\rm orb}(\vec r) &=& V^{\rm orb}(r)\sum_{i=a,b} \frac{3(\hat r \cdot \vec \jmath_i)(\hat r \cdot \vec \jmath_i)-\vec \jmath_i \cdot \vec \jmath_i  }{\sqrt{6}}+ \cdots \,, \nonumber
\end{eqnarray}
and 
\begin{eqnarray*}
U^{\rm dip}(\vec r)&=&- \frac{\mu_0(g\mu_{\rm B})^2}{4\pi}  \frac{3(\hat r \cdot \vec \jmath_a)(\hat r \cdot \vec \jmath_b)-\vec \jmath_a \cdot \vec \jmath_b}{r^3}
\end{eqnarray*}
is the  magnetic dipole-dipole interaction.  We use $ V^{\rm
orb}(r)=-c^{\rm orb}_6/r^6 $ for all $r$  with $c^{\rm orb}_6 =
-1.904E_{\rm h} a^6_0$.  Finally, $E_{\rm h}$ is the Hartree energy,
$c$ is the speed of light in vacuum, and $\mu_0$ is the magnetic constant.

The coefficient $C_6$ is the largest dispersion coefficient by far, making
the van-der-Waals length $R_6=\sqrt[4]{2\mu C_6/\hbar^2}$  the natural
length scale for the dispersive interactions. The contribution to $U^{\rm
orb}(\vec r)$ with strength $V^{\rm orb}(r)$ is the strongest anisotropic
orbital interaction. It couples the angular momentum of each atom  to the
rotation of the molecule.  For future reference and following the convention
in the literature the natural length scale for the magnetic dipole-dipole
interaction is  $a_{\rm dd}=(1/3)\times 2\mu C_3/\hbar^2$ with coefficient
$C_3=\mu_0(g\mu_{\rm B}j)^2/(4\pi)$, where $j=6$ and $g=1.16381$ for Er.

The Hamiltonian $H_{\rm rel}$ commutes with  $J_{z}$ and only  couples even
or odd  values of $\vec \ell$, the relative orbital angular momentum or
partial wave.  Here, $J_{z}$ is the $z$ projection of the total angular
momentum $\vec J =\vec \ell + \vec \jmath\,$ with $\vec \jmath=\vec \jmath_a
+ \vec \jmath_b$.  For $B=0$ $H_{\rm rel}$ also commutes with $J ^2$.
Eigenstates $| i, M \rangle$ of $H_{\rm rel}$ with energy $E_{i,M}$
are labeled by projection quantum number $M$ and index $i$. These eigen pairs  have been
computed  in the spin basis \[ |(j_aj_b)j \ell; J  M \rangle\equiv
    \sum_{m_j m}   \langle j\ell m_j m | J  M  \rangle
    |(j_aj_b)j m_j\rangle Y_{\ell m}(\theta,\varphi) \] with $|(j_aj_b)j
m_j\rangle=\sum_{m_a m_b}\langle j_a j_b m_a m_b | j m_j \rangle |j_a
m_a\rangle |j_b m_b\rangle$, spherical harmonic function $Y_{\ell
m}(\theta,\varphi)$, and  $\langle j_1 j_2 m_1 m_2 | j m \rangle$ are
Clebsch-Gordan coefficients.  For our bosonic system only basis states with
even $\ell+j$ exist.  We use a discrete variable representation (DVR)
\cite{ColbertMiller,Tiesinga1998} to represent the radial coordinate $r$.
The largest $r$ value is  a few times the largest of the harmonic oscillator
lengths $\sqrt{\hbar/(\mu\omega_{\rho,z})}$  and for
typical traps $R_6\ll \sqrt{\hbar/(\mu\omega_{\rho,z})}$.  We further
characterize eigen states by computing overlap amplitudes with those at
different $B$ field values. In particular, overlaps with $B=0$ eigenstates
give us the approximate $J$ value.  The expectation value of operators $j^2$ and
$\ell^2$ are also computed.

\subsection*{Eigenstates in an isotropic lattice site}

\begin{figure}
\includegraphics[scale=0.15,trim=0 0 0 0,clip]{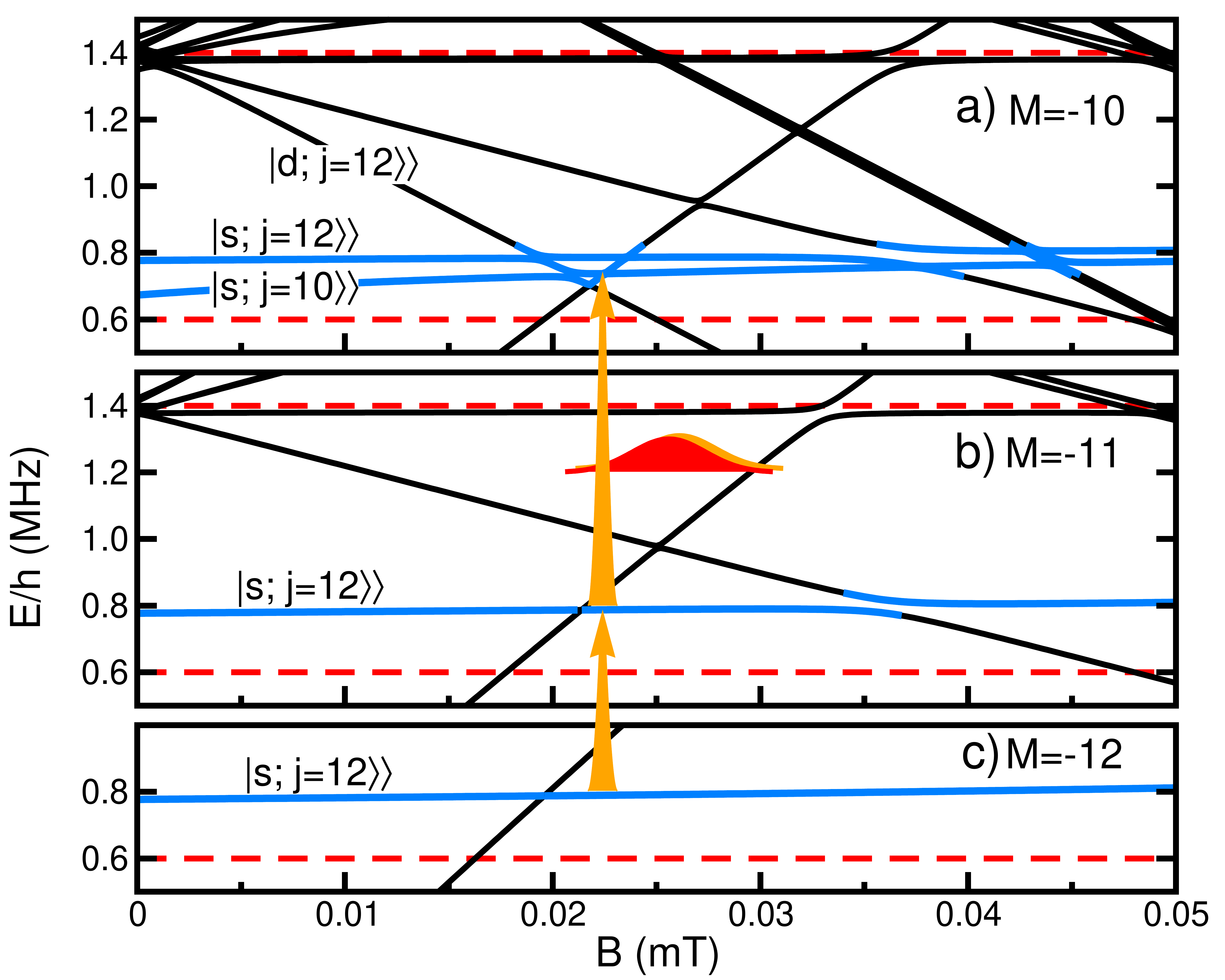}
\caption{Near-threshold energy levels of the relative motion of two
harmonically trapped and interacting $^{168}$Er atoms with ${M =-10}$,
$-11$, and $-12$ in  panel a), b), and c), respectively.  The  trap is
isotropic with $\omega/(2\pi)$ = 0.4 MHz.  In the three panels the zero
of energy  corresponds to the Zeeman energy of two atoms at rest with
${m_a+m_b=-10}$, $-11$, and $-12$, respectively.    Dashed red lines
correspond to the first and second harmonic-oscillator
levels with energies $(3/2)\hbar\omega$ and $(7/2)\hbar\omega$,
respectively.  Blue curves indicate the energy levels that are involved
in the spin-changing oscillations. Some of the eigenstates have been
labeled by their dominant  $\ell$ and  $j$ contribution.  Orange arrows
(not to scale) and a red rf pulse indicate single-color two-photon rf
transitions that initiate the spin oscillations starting from two atoms
in the $|6,-6\rangle$ state.
 }
\label{IsoTrap}
\end{figure}

Figure~\ref{IsoTrap} shows even-$\ell$, above-threshold $^{168}$Er$_2$
energy levels in an isotropic harmonic trap  and $M=-12$, $-11$, or $-10$ as
functions of  $B$ up to  0.05 mT.   The energies of two harmonic oscillator
levels are also shown.  The energetically-lowest is an $\ell=0$ or $s$-wave
state with energy $(3/2)\hbar \omega$; the second is degenerate with one
$s$- and multiple $\ell=2$, $d$-wave states and energy $(7/2)\hbar \omega$.  In
each panel one or two eigenstates with energies that run nearly parallel
with these  oscillator energies exist. In fact, their energy, just above
$(3/2)\hbar \omega$, indicates a repulsive effective atom-atom interaction
\cite{Busch1998,Tiesinga2000}.  For our weak $B$ fields and {\it away} from
avoided crossings their wavefunctions are well described by or are
correlated to a single ${J =10}$ or 12 zero-$B$ eigenstate. Further, they
have an $s$-wave dominated spatial function,  ${j\approx J }$, and
${m_j\approx M }$.  These eigenstates will be labeled by $| s; j m_j
\rangle\rangle$ away from avoided crossings.

A single bound state with $E<0$ when $B=0$ and a negative magnetic
moment can be inferred in each panel of Fig.~\ref{IsoTrap}. It avoids
with several trap states when $B>0.015$ mT and has  mixed $g$- and $i$-wave  character
away from avoided crossings.   In free space  this bound state would
induce a narrow Feshbach resonance near $B=6$ $\mu$T (not shown).

Figure~\ref{IsoTrap} also shows  eigenstates with energies close
to $E=(7/2)\hbar \omega$ when $B=0$. For $B>2$ $\mu$T  and away from
avoided crossings, these states have  $d$-wave character, are well described by
a single $j,m_j$ pair, and have a magnetic moment, $-dE_{i,M}/dB$, that
is an integer multiple of $g\mu_{\rm B}$. 
$D$-wave states with a positive magnetic moment  in the figure have
avoided crossings with $s$-wave  states $| s; j m_j \rangle\rangle$
and play an important role in our analysis of spin oscillations.
We focus on the three-state avoided crossing in panel a) near ${B=
0.020}$ mT, where the corresponding $d$-wave state has $j=12$ and
$m_j=-12$ and will be labeled by $| d; j m_j\rangle\rangle$.  Close to 
the avoided crossings the three eigenstates of $H_{\rm rel}$
are  superpositions of the $B$-independent $| s; j m_j \rangle\rangle$
and $| d; j m_j\rangle\rangle$ states. The mixing coefficients follow
from the overlap amplitudes with eigenstates well away from the avoided
crossing. In other words $ | i,  {M =-10}\rangle= \sum_k U_{i,k}(B)\, | k
\rangle\rangle $ with $k=\{s;10,-10\}$, $\{s;12,-10\}$, and $\{d;12,-12\}$
and  $U(B)$ is a $B$-dependent  $3\times 3$ unitary matrix.

Finally, Fig.~\ref{IsoTrap} shows the rf pulse that initiates spin
oscillations. We choose a non-zero $B$ field and start in 
$|s; 12\,{-12}\rangle\rangle$ with  ${M =-12}$ in panel c).
After the pulse, via near-resonant intermediate states  with ${M
=-11}$, the  atom pair is in a superposition of  two or three eigenstates
with  ${M =-10}$. The precise superposition depends on experimental
details such as carrier frequency, polarization
and pulse shape.  We, however, can use the following observations.
The initial $|s; 12\,{-12}\rangle\rangle$ state can also be expressed
as  the uncoupled product state $|6,-6\rangle|6,-6\rangle$ independent
of $B$. As rf photons only induce transitions in  atoms (and not couple
to $\vec \ell$ of the atom pair), the absorption of one photon by each
atom  leads to the product $s$-wave state $|6,-5\rangle|6,-5\rangle$,
neglecting changes to the spatial wavefunction of the atoms. To prevent
population in atomic Zeeman levels with ${m_{a,b}>-5}$ we  follow
Ref.~\cite{Laburthe2013} and assume that  light shifts induced by optical
photons briefly break the resonant condition to such states.

The  $s$-wave  $|6,-5\rangle|6,-5\rangle$ state then evolves under the
molecular Hamiltonian. It is therefore convenient to express this state
in terms of ${M =-10}$ eigenstates.  First,  by coupling the atomic
spins to $\vec \jmath$ we note $|6,-5\rangle|6,-5\rangle \to c_{10}
| s; 10,{-10}\rangle\rangle + c_{12}|s; 12,{-10}\rangle\rangle$,
where $c_j=\langle 6 6 \,{-5} {-5} | j \,{-10}\rangle$ and each
$|k\rangle\rangle$ is a superposition of three eigenstates $| i, {M
=-10}\rangle$ as given by the inverse of $U(B)$.  
After free evolution for time $t$  we measure the population
of remaining in the initial state.

\subsection*{Anisotropic harmonic traps}

The lattice laser beams along the independent spatial directions do not
need to have the same intensity and thus the potential for a lattice site
can be anisotropic. This gives us a means to extract further information
about the anisotropic molecular interaction potentials.  A first such
experiment for Er provided evidence of the orientational dependence
due to the intra-site magnetic dipole-dipole interactions
on the particle-hole excitation frequency in the doubly-occupied Mott
state \cite{Baier2016}.

\begin{figure}
	\includegraphics[scale=0.27,trim=0 0 10 80,clip]{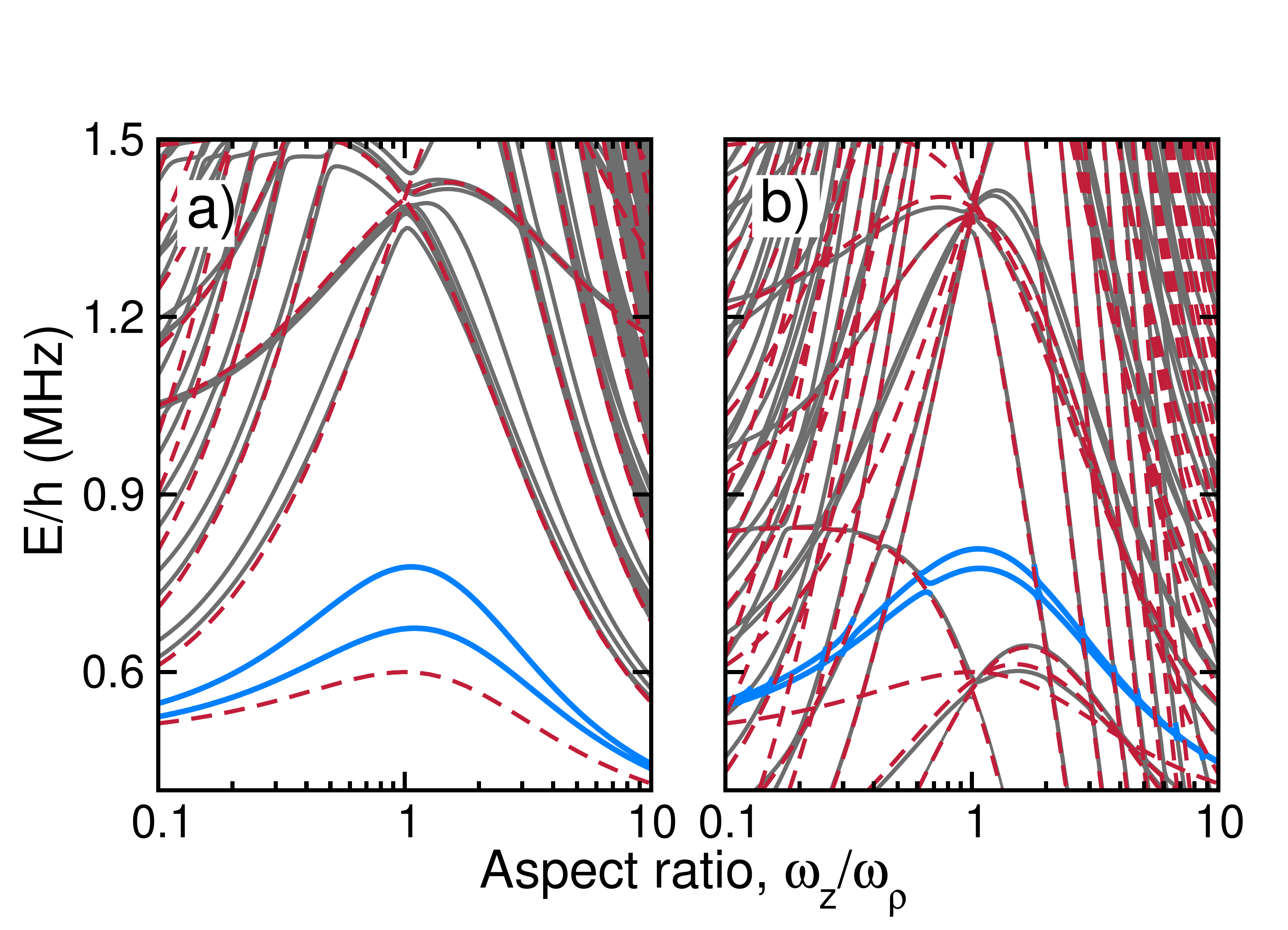}
	\caption{Near-threshold energy levels of $^{168}$Er$_2$ with ${M =-10}$ in an anisotropic
harmonic trap as a function of trap aspect ratio $\omega_z/\omega_\rho$
at fixed $\omega/(2\pi)=  0.4$ MHz for  $B=0.1$ $\mu$T and 0.05 mT
in panels a) and b), respectively. The red dashed lines correspond
to non-interacting levels.  Blue lines are relevant for
spin-oscillation experiments.  The zero of energy in a panel is that of
two free atoms at rest with $m_a+m_b=-10$ at the corresponding $B$ field.
} \label{AnisoTrap} \end{figure}

Figure \ref{AnisoTrap} shows the near-threshold energy ${M
=-10}$ levels in an anisotropic harmonic trap as a function of trap aspect
ratio $\omega_z/\omega_\rho$ at fixed $\omega/(2\pi)=  0.4$ MHz for two
magnetic field strengths. Dashed lines  correspond to  non-interacting
levels, including only the harmonic trap and Zeeman energies.  For 
$\omega_z/\omega_\rho\to 0$ and $\infty$
the trap is quasi-1D and quasi-2D, respectively.  Energy levels 
involved in spin-oscillations are highlighted in blue and for finite $B$ their
avoided crossings with other states can be studied.

\bibliography{Refs}